\documentclass[epj]{svjour}
\usepackage{amsmath}
\usepackage{graphics}
\usepackage{dcolumn}
\usepackage{bm}

\newcommand{\lw}[1]{\smash{\lower 1.5ex\hbox{#1}}}
\begin{document}
\title{Extended Gaussian ensemble or $q$-statistics\\
       in hadronic production processes?}
\author{T.Osada\inst{1}\thanks{\emph{e-mail:}
osada@ph.ns.musashi-tech.ac.jp} \and
O.V.Utyuzh\inst{2}\thanks{\emph{e-mail:} utyuz@fuw.edu.pl} \and
G.Wilk\inst{2}\thanks{\emph{e-mail:} wilk@fuw.edu.pl} \and Z.W\l
odarczyk\inst{3}\thanks{\emph{e-mail:} wlod@pu.kielce.pl}
}                     
%
%
\institute{Dept.of Physics, General Education Center, Musashi Institute
of Technology, Tamazutsumi 1-28-1, Setagaya-ku, Tokyo 158-8557, Japan
\and The Andrzej So\l tan Institute for Nuclear Studies, Ho\.za 69, 00681
Warsaw, Poland\and Institute of Physics, \'Swi\c{e}tokrzyska Academy,
\'Swi\c{e}tokrzyska 15; 25-406 Kielce, Poland}
\date{Received: date / Revised version: date}
%
\abstract{ The extended Gaussian ensemble introduced recently as a
generalization of the canonical ensemble, which allows to treat energy
fluctuations present in the system, is used to analyze the inelasticity
distributions in high energy multiparticle production processes.
} 
\maketitle

%
%

\section{Introduction} \label{intro}

 The high energy multiparticle production processes are very
important source of information on the dynamics of hadronization process,
in which some amount of the initially available energy is subsequently
transformed into a number of secondaries of different types. Such
processes can be described only via phenomenological models, which are
stressing their different dynamical aspects, like specific energy flows
\cite{IGM} or their apparent thermal-like character \cite{stat}. Actually
most of the characteristic features of hadronization can be described in
universal manner by means of Information Theory (IT) approach, both in
its extensive \cite{MaxEnt} or nonextensive \cite{MaxEntq,PHYSA,Qrest}
versions. The main difference between them is that whereas former is
using only energy-momentum conservation constraint, the later accounts
also for some intrinsic fluctuations present in the hadronization
process, either in the form of fluctuations of temperature \cite{WW} or
in the form of fluctuations of the number of produced secondaries
\cite{MaxEntq}\footnote{Accounting for the fact that multiplicity
distribution of observed secondaries are not Poissonian \cite{NBD}.}.
Recently the extended gaussian ensemble (EGE) approach has been proposed
to account for some fluctuations in statistical mechanics and it was
presented also in the IT formulation \cite{eGe}. The question, which we
would like to address here, is whether EGE can find application in
deducing some new information from hadronic production processes.

\section{Extended Gaussian ensemble from IT}

Following \cite{MaxEnt,MaxEntq,PHYSA,Qrest} we are interested in applying
IT to deduce the most probable and least biased energy distributions of
particles produced in hadronization process in which mass $M$ transforms
into given number $N$ of secondaries of mass $\mu$ and mean transverse
mass $\mu_T = \sqrt{\mu^2 +\langle p_T\rangle^2}$ each, distributed in
the longitudinal phase space described by rapidity variable, $y$ (such
that energy of particle is $E=\mu_T\cosh y$). We are therefore interested
in (normalized) rapidity distribution $p(y) = (1/N)\cdot dN/dy$, $\int dy
p(y) = 1$, which according to IT \cite{MaxEnt} is obtained by maximizing
Shannon entropy
\begin{equation}
S = - \int dy p(y) \ln p(y), \label{eq:Shannon}
\end{equation}
under condition of reproducing known {\it a priori} mean value of energy
of produced secondaries ($K$ denotes the so called {\it inelasticity} of
reaction to be discussed later),
\begin{equation}
\langle E(y)\rangle = \int_{-Y_m}^{Y_m} dy \left[\mu_T\cdot \cosh
y\right]\cdot p(y) = U = \frac{K}{N} \cdot M . \label{eq:inel}
\end{equation}
Whereas in \cite{MaxEntq} one uses Tsallis entropy instead Shannon ones
and defines constraints (\ref{eq:inel}) in slightly different way, the
EGE approach \cite{eGe} simply adds one more constraint to
(\ref{eq:inel}) in the form of {\it a priori} known fluctuations of mean
energy of given secondary given by its variance $W$,
\begin{equation}
\langle [E(y) - U]^2\rangle = \int_{-Y_m}^{Y_m} dy \left[\mu_T\cdot \cosh
y - U\right]^2\cdot p(y) = W .
\label{eq:fluct}
\end{equation}
In this case \cite{eGe}
\begin{equation}
p(y) = \frac{1}{Z}\cdot \exp\left[ - \beta\cdot \mu_T\cosh y -
\gamma\cdot \left(U - \mu_T\cosh y\right)^2\right], \label{eq:p(y)g}
\end{equation}
where $Z$ is normalization constant and $\beta=\beta(U,W,N,\mu_T)$,
$\gamma=\gamma(U,W,N,\mu_T)$ are two Lagrange multipliers for the
constraints (\ref{eq:inel}) and (\ref{eq:fluct}), respectively. In the
case of no dynamical fluctuations, i.e., $\gamma =0$, one recovers
situation already known from \cite{MaxEnt,MaxEntq} (with some $W_0 =
1/\beta^2$ with respect to which one should estimate effect of dynamical
correlations)\footnote{In the center of mass frame $y \in \left(-Y_m,
Y_m\right)$ where $Y_m = \ln \left[ M' \left(1+\sqrt{ 1-
4\mu_T^2/M'^2}\right)/(2\mu_T)\right]$ and where $M' =M - (N-2)\mu_T$.}.
Rewriting eq. (\ref{eq:p(y)g}) as
\begin{equation}
p(y) = \frac{1}{Z}\cdot \exp\left( - \beta^{*}\cdot \mu_T\cosh y \right)
;  \beta^{*} = \beta - \gamma\cdot \left[ 2U - E(y)\right],
 \label{eq:p(y)q}
\end{equation}
one obtains expression formally resembling the usual Boltz\-mann-Gibbs
formula, but this time with energy-dependent inverse "temperature"
$\beta^{*}$ (which is thus no longer intensive variable). Actually such
possibility was already discussed in \cite{Almeida} in the context of
reservoir with finite heat capacity. It was argued there that if
\begin{equation}
\frac{d}{dE}\left[\frac{1}{\beta(E)}\right] = q-1, \label{eq:almeida}
\end{equation}
where $q$ is some constant, then the corresponding distribution (where
$E(y) = \mu_T\cosh y$) takes form of the so called Tsallis distribution
\cite{CT},
\begin{equation}
p_q(y) \,=\, \frac{1}{Z_q(M,N)} \left[ 1\, -\, (1 - q) \beta_q(M,N)\cdot
E(y)\right]^{\frac{1}{1-q}}, \label{eq:p(y)q}
\end{equation}
with $q$ given by (\ref{eq:almeida}) \footnote{Care must be taken here
when considering signs because in \cite{Almeida} one considers dependence
of $\beta$ on the energy of the reservoir, $E_R$, and here we have energy
of particle $E=E_{total} - E_R$. Therefore our $q-1$ corresponds to $1-q$
there.}. In our case where $\beta = \beta^{*}$ one gets {\it formally}
energy dependent Tsallis nonextensivity $q$ parameter
\begin{equation}
q = 1 - \frac{\gamma}{[(\beta - 2\gamma U) + \gamma E]^2}.
\label{eq:fullq}
\end{equation}
For $\gamma > 0$ it becomes smaller than unity and exceeds unity for
$\gamma < 0$. It coincides with result of \cite{eGe} only if $|\gamma
(E-2U)/\beta |<<1$ in which case $q=1 - \gamma/\beta^2$.

\section{Inelasticity distributions $\chi(K)$ in EGE}

\begin{figure}
  \begin{center}
  \resizebox{0.41\textwidth}{!}{\includegraphics{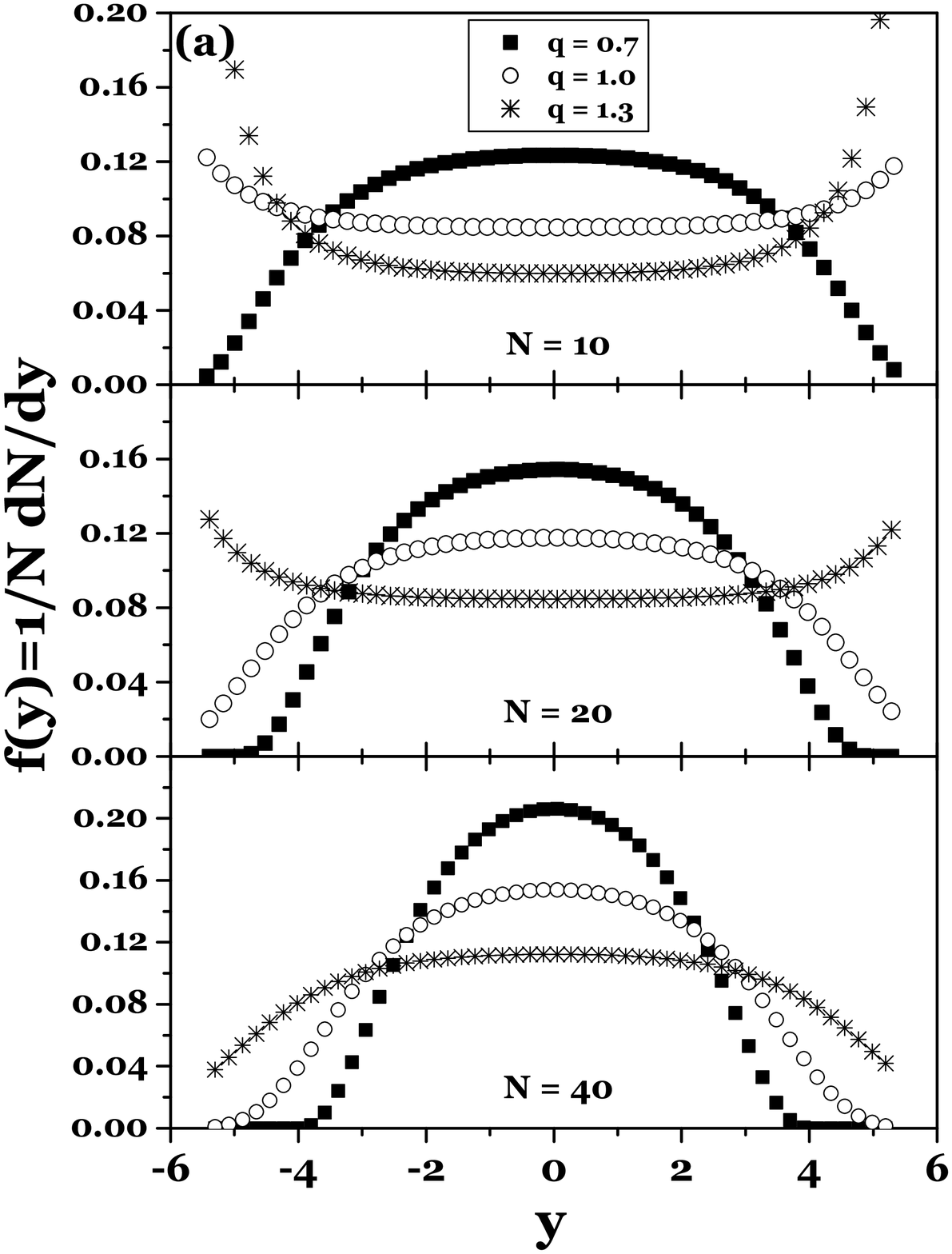}}
  \resizebox{0.41\textwidth}{!}{\includegraphics{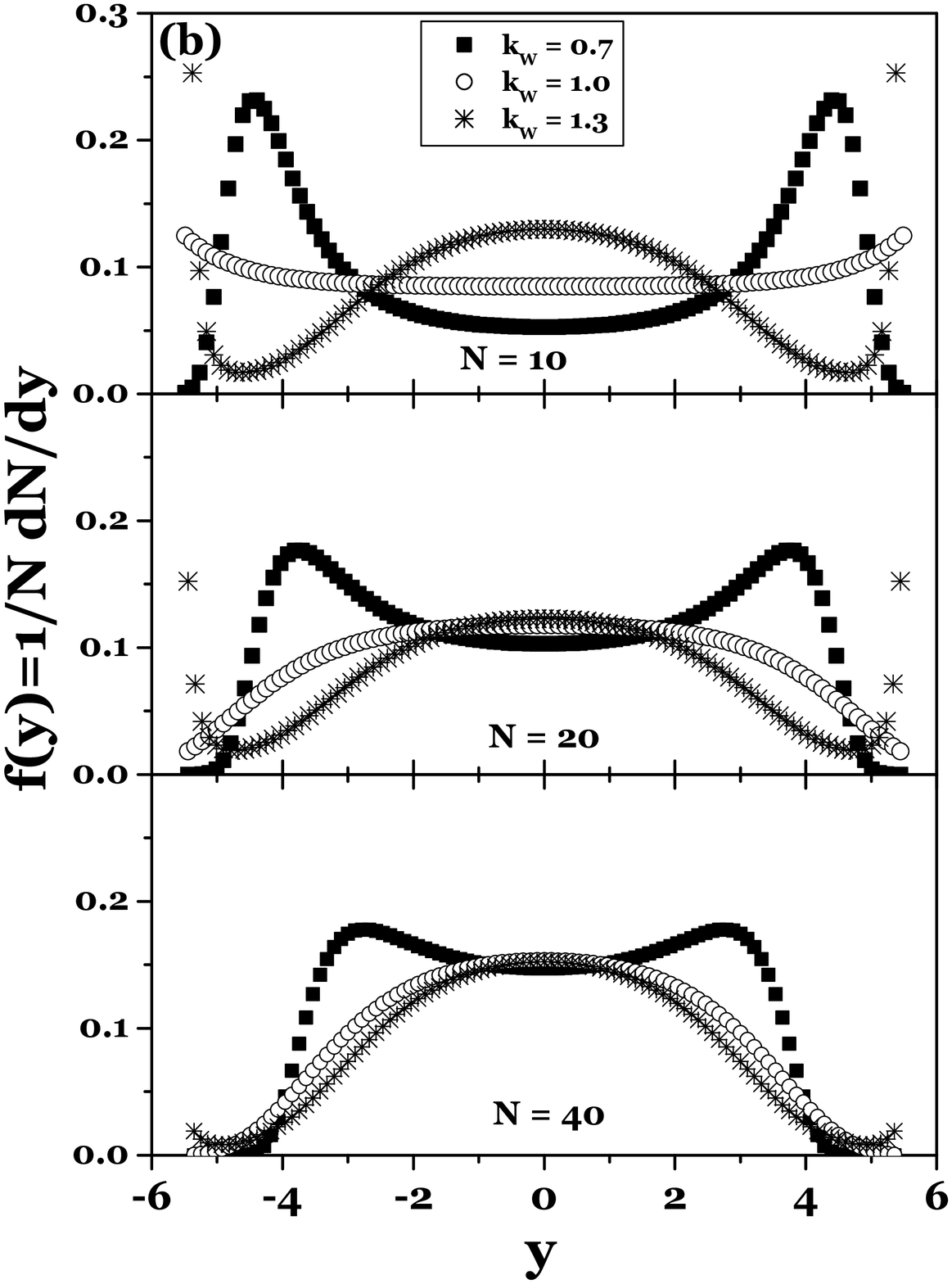}}
  \end{center}
  \caption{$(a)$ Examples of the most probable rapidity distributions
  as given by eq. (\ref{eq:p(y)q}) for hadronizing mass $M=100$ GeV decaying
  into $N$ secondaries of (transverse) mass $m_T=0.4$ GeV each for different
  values of $q$ parameter.
  $(b)$ The same but using eq. (\ref{eq:p(y)g}) with different values of
  fluctuations given by $k_W$ such that $W=k_W^2W_0$ where $W_0$ are the intrinsic
  statistical fluctuations present in the system when $\gamma=0$, i.e., where
  gaussian becomes pure exponential.}
  \label{fig:pqg}
\end{figure}

We have tried to apply EGE distribution as defined by eq.
(\ref{eq:p(y)g}) to analyze the same multiparticle data as in
\cite{MaxEntq} only to discover that these data do not require EGE, the
best fit is obtained with $\gamma =0$ or slightly negative (in which case
the respective $q$ from (\ref{eq:fullq}) exceeds unity, as has been found
in \cite{MaxEntq}). The reason for this is obvious when inspecting Fig.
\ref{fig:pqg},
which confronts rapidity distributions of the Tsallis type
(\ref{eq:p(y)q}) with those obtained from EGE (\ref{eq:p(y)g}) obtained
for hadronization of some fixed mass $M$ into different number of
secondaries. Results obtained using EGE show completely different
behavior from Tsallis statistics approach clearly demonstrating that
direct fluctuations in energy used in EGE (and characterized here by
parameter $k_W$ such that $W=k_W^2W_0$ where $W_0$ are the intrinsic
statistical fluctuations present in the system when $\gamma=0$) {\it are
not equivalent} to fluctuations described by parameter $q$ of Tsallis'
statistics\footnote{ Notice that for $k_W<1$ one gets $\gamma >0$
(actually $\gamma \rightarrow +\infty$ for $k_W \rightarrow 0$) whereas
for $k_W > 1$ one obtains $\gamma <0$ leading to equivalent $q$
calculated according to eq. (\ref{eq:fullq}) exceeding unity but
otherwise being uncompatible with nonextensivity parameter used in upper
panel of Fig. \ref{fig:pqg}.}. This can be understood in the following
way. In standard description of hadronization processes by means of IT in
the Shannon form we always have some (mean) number of secondaries
produced $\langle N\rangle$ with (mean) energy $\langle E\rangle$ each.
Allowing for fluctuations of $\langle N\rangle $ results in
$q$-statistics using Tsallis entropy for IT \cite{MaxEntq}. In this case
the mean energy per particle fluctuates from event to event. Keeping now
$\langle N\rangle$ fixed but introducing {\it distribution} of energy per
particle (i.e., describing energy per particle by its mean and deviation
from the mean) results in EGE\footnote{Notice that decreasing
fluctuations in energy in comparison to standard ones, i.e., assuming in
Fig. \ref{fig:pqg} (lower panel) $k_W<1$ results in tendency of particles
to condensate in a single energy state with energy equal to
$E_{total}/N_{total}$. It means that EGE interpolates in fact between the
microcanonical and canonical distributions \cite{interpol}.}. Evidently
single particle distributions in hadronization processes follow first or
second scenario, not EGE.
\begin{figure}
  \begin{center}
  \resizebox{0.48\textwidth}{!}{\includegraphics{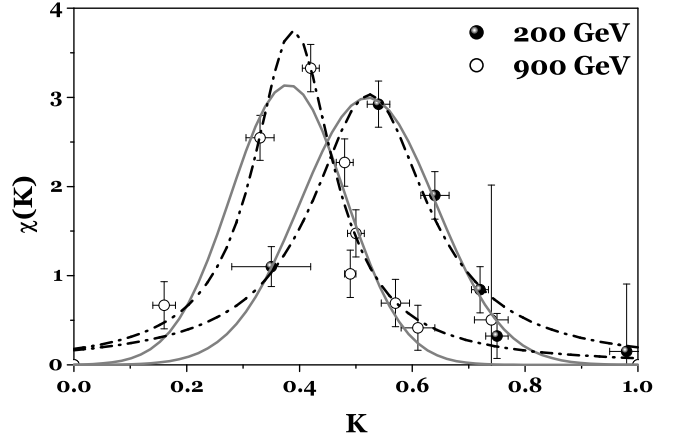}}
  \end{center}
  \caption{Inelasticity distributions $\chi(K)$ (normalized to
unity) obtained in \cite{MaxEntq} from analysis of multiparticle
production data for $\sqrt{s} = 200$ GeV and $\sqrt{s} = 900$ GeV fitted
by gaussian, $\chi(K) \simeq \exp \left[ -(K-\langle
K\rangle)^2/(2\sigma^2) \right]$ (full lines), and lorentzian, $\chi(K)
\simeq \sigma/[4(K-\langle K\rangle)^2 + \sigma^2]$ (dash-dotted lines)
formulas, respectively. The values of parameters $(<K>;\sigma)$ for
gaussian case are $(0.52;0.24)$ and $(0.38;0.20)$ for $200$ GeV and $900$
GeV, respectively; for lorentzian case they are, respectively,
$(0.52;0.25)$ and $(0.39;0.17)$ (see \cite{MaxEntq} for more details).
            }
  \label{fig:ChiK}
\end{figure}

On the other hand EGE turns out to be very useful when applied to other
characteristic of multiparticle production, namely to inelasticity
distribution, $\chi(K)$, (i.e., distribution of the fraction of the
available energy, which is transformed into observed secondaries). In
\cite{MaxEntq} it was deduced from data for the first time for two
energies: $200$ and $900$ GeV, cf. Fig. \ref{fig:ChiK} (by analysing
rapidity distributions of secondaries in fixed multiplicity bins). Its
shape has been then fitted by gaussian and lorentzian curves but no
explanation was offered for their possible origin and there was no
argument at that time in favor of any of them. EGE provides arguments
that most probably $\chi(K)$ should be of gaussian shape. To show this
let us again follow \cite{eGe} and let us suppose that the whole energy
available for a given multiparticle production reaction, $E=\sqrt{s}$, is
divided into two parts: one part equal to $E_1=K\cdot\sqrt{s}$ is going
into system producing observed secondaries whereas the rest of it, $E_2 =
E-E_1$, is not used for this purpose and, in a sense, acts as a kind of
"heath bath" (or environment) for the first one. Both systems, the one
producing particles with energy $E_1$ and the environment with energy
$E_2$ can be in many possible states. Therefore
\begin{equation}
p_1(E_1) = \frac{\Omega_1(E_1)\Omega_2(E_2)}{\Omega_{1+2}(E)},
\label{eq:Omega12}
\end{equation}
where $\Omega$ denote the corresponding number of states. Defining
entropy in the usual way as
\begin{equation}
S_i(E_i) = \ln \Omega_i(E_i),\qquad i=1,2  \label{eq:S}
\end{equation}
one gets
\begin{equation}
p_1(E_1) = \frac{1}{\Omega_{1+2}(E)}\cdot
\exp\left[S_1(E_1)+S_2(E_2)\right]. \label{eq:P(S)}
\end{equation}
\begin{figure}
  \begin{center}
  \resizebox{0.44\textwidth}{!}{\includegraphics{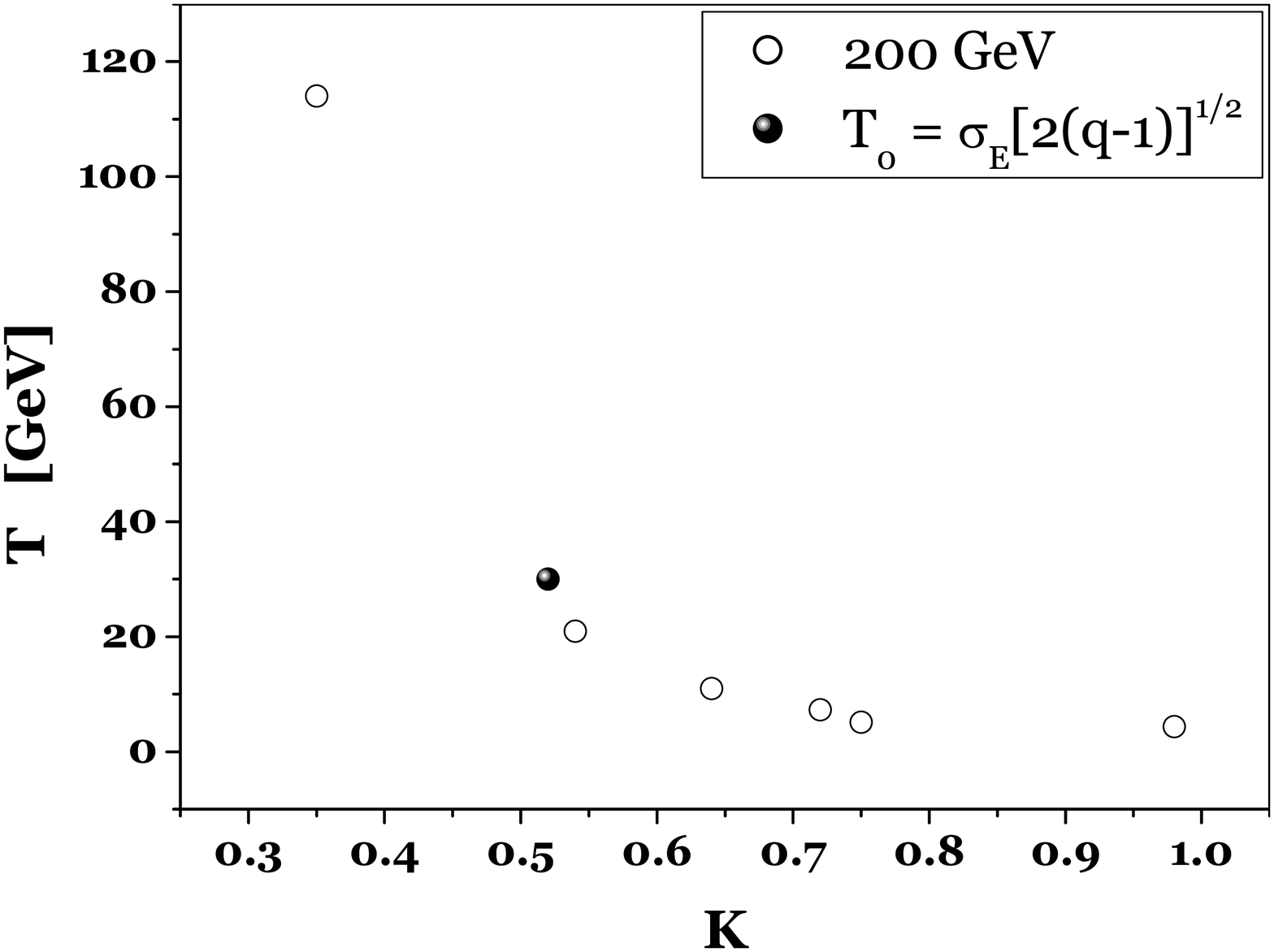}}
  \resizebox{0.45\textwidth}{!}{\includegraphics{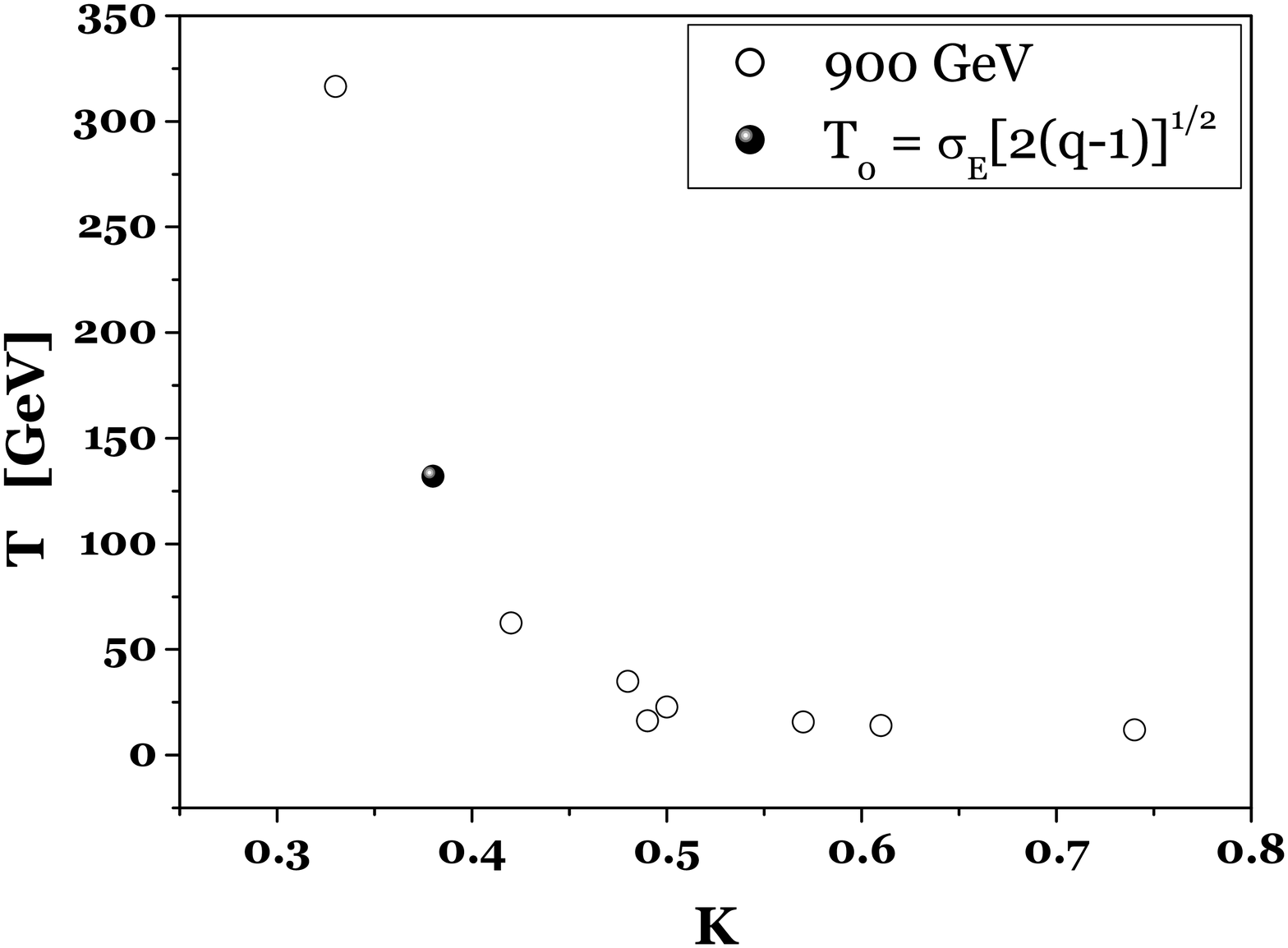}}
  \end{center}
  \caption{Inelasticity dependence of the temperature parameter as
deduced in \cite{MaxEntq} from multiparticle data at energies $200$ and
$900$ GeV. Black circles correspond to the corresponding temperature
obtained from the thermodynamical analysis performed along the EGE
ideas.}
  \label{fig:T}
\end{figure}
Expanding now entropy around $E_1=U$, keeping only linear and quadratic
terms and assuming that $\beta = \frac{1}{T_0} = \left[\frac{\partial \ln
\Omega} {\partial E_1} \right]_{E_1=U}$ and $\gamma = -
\left[\frac{\partial^2\ln \Omega} {\partial E_1^2}\right]$ are the same
for both parts of the system (generalization is straightforward) one
immediately obtains gaussian-like form for energy $E_1$ distribution,
\begin{equation}
p_1(E_1) = \frac{1}{Z_G}\exp\left[  - \gamma\left(E_1 -
U\right)^2\right],\label{eq:PeGe}
\end{equation}
which, because $E_1 = \sqrt{s}\cdot K$ and $U=\sqrt{s}\cdot \langle
K\rangle$ (where $\sqrt{s}$ is energy of reaction), translates in natural
way to gaussian distribution of inelasticity, $\chi(K)$, as the most
probably form with $\sigma^2 =\frac{1}{2\gamma}$. From \cite{MaxEntq} one
can also deduce the $K$ dependence of the temperature $T$. Notice that in
our case parameter $\gamma$ can be connected with $T$ and heat capacity
$C_V$ because $\gamma = 1/(C_VT^2_0)$. On the other hand \cite{MaxEntq}
provides us also with parameter $q$ for different inelasticities for both
energies and we know that $C_V$ can be connected with the nonextensivity
parameter $q$, namely $1/C_V=q-1$ \cite{WW}\footnote{Notice, however,
that now $C_V$ and $q$ are different from those in \cite{WW} as they
refer to both longitudinal and transverse degrees of freedom, cf.
\cite{PHYSA} for discussion how $q_L$ and $q_T$ can be composed to
produce total $q$.}. It means therefore that there is simple relation
connecting the width $\sigma$ of the observed gaussian distribution
(i.e., parameter $\gamma$ of EGE), the temperature $T_0$ and the
nonextensivity parameter $q$ describing internal behavior of the selected
subsystem, namely
\begin{equation}
T_0 = \sqrt{2(q-1)}\sigma_E\qquad {\rm where}\qquad \sigma_E=\sigma\cdot
\sqrt{s} .\label{eq:T0}
\end{equation}
As can be seen in Fig. \ref{fig:T}, $T_0$ deduced in such way agrees very
well with the $(T,K)$ dependence deduced from analysis of rapidity
distributions in fixed multiplicity bins \cite{MaxEntq} .

\section{Conclusions}

We would like to conclude with the following remarks:
\begin{itemize}
\item EGE works {\it only} for the whole system, not for a single
particle. This is going to be emitted according to its own distribution,
in particular Boltzmann-Gibbs or Tsallis, notwithstanding what the energy
$E_1$ is and how it is distributed. For a moment we cannot offer any
convincing explanation why it is so.

\item EGE is not the same (i.e., it does not describes the same kind of
fluctuations) as $q$-statistics. It means that even if for some limiting
cases both distribution can be similar this is just an artifact.

\item On the other hand EGE tells us that for the system under
consideration $T=T(E)$ and $E_1$ fluctuates. This means that for
particles emitted from this system one should rather use Tsallis
distributions reserving Boltzmann-Gibbs ones only to the case of
$T$=const.
\end{itemize}

Let us close with remark that the lorentzian curve shown also in Fig.
\ref{fig:ChiK} (and fitting data at least as well as the gaussian one)
could be explained as a kind of a nonextensive extension of EGE by
noticing that in q-statistical approach one gets gaussian distribution
for $q=1$ and lorentzian distribution for $q=2$. We shall not pursue this
further here.

\noindent {\bf Acknowledgments:} GW is grateful to the organizers of the
{\it NEXT2005} for their hospitality. Partial support of the Polish State
Committee for Scientific Research (KBN) (grant
621/E-78/SPUB/CERN/P-03/DZ4/99(GW)) is acknowledged.


\begin{thebibliography}{}

\bibitem{IGM}
F.~O.~Dur\~aes, F.~S.~Navarra and G.~Wilk, Braz.\ J.\ Phys.\ \textbf{35},
3 (2005).

\bibitem{stat}
Cf., for example, F.~Becattini, Nucl.\ Phys.\ A \textbf{702}, 336 (2002);
F.~Becattini and G.~Passaleva, Eur.\ Phys.\ J.\ C \textbf{23}, 551
(2002); W.~Broniowski and W.~Florkowski, Phys.\ Rev.\ Lett. \textbf{87},
272302 (2001) and Acta Phys.\ Polon.\ B \textbf{35}, 779 (2004) and
references therein.

\bibitem{MaxEnt}
G.~Wilk and Z.~W\l odarczyk, Phys.\ Rev.\  D \textbf{43}, 794 (1991).

\bibitem{MaxEntq}
F.~S.~Navarra, O.~V.~Utyuzh, G.~Wilk and Z.~W\l odarczyk, Phys.\ Rev.\ D
\textbf{67}, 114002 (2003).

\bibitem{PHYSA}
F.~S.~Navarra, O.~V.~Utyuzh, G.~Wilk and Z.~W\l odarczyk, Physica A
\textbf{340}, 467 (2004).

\bibitem{Qrest}
F.~S.~Navarra, O.~V.~Utyuzh, G.~Wilk and Z.~W\l odarczyk, Physica A
\textbf{344}, 568 (2004) and Nukleonika \textbf{49} (Supplement
\textbf{2}), s19 (2004), see also O.~V.~Utyuzh, G.~Wilk and Z.~W\l
odarczyk, {\it Multiparticle production processes from the Information
Theory point of view}, hep-ph/0503048, to be published in Acta Phys.\
Hung.\ (HIP) (2005).

\bibitem{WW}
G.~Wilk and Z.~Wlodarczyk, Phys.\ Rev.\ Lett.\ \textbf{84}, 2770 (2000);
Chaos,\ Solitons\ and\ Fractals {\bf 13/3}, (2001) 581; Physica A
\textbf{305}, 227 (2002). Cf. also: C.~Beck and E.~G.~D.Cohen, Physica A
\textbf{322}, 267 (2003) and T.~S.~Bir\'o and A.~Jakov\'ac, Phys.\ Rev.\
Lett. \ \textbf{94}, 132302 (2005).

\bibitem{NBD}
A.~Giovannini and L.~Van Hove, Z.\ Phys.\ C \textbf{30}, 381(1986) 381;
see also: P.~Carruthers and C.~S.~Shih, Int.\ J.\ Mod.\ Phys.\  A
\textbf{2}, 1447 (1986).

\bibitem{CT}
C.~Tsallis, Physica A \textbf{340}, 1 (2004) and Physica A \textbf{344},
718 (2004) and references therein.

\bibitem{eGe} R.S.Johal, A.Planes and E.Vives, {\sl Phys. Rev.} {\bf E68} (2003)
              056113.

\bibitem{Almeida}
 M.~P.~Almeida, Physica A \textbf{300}, 424 (2001).

\bibitem{interpol}
M.~S.~S.~Challa and J.~H.~Hetherington, Phys.\ Rev.\ Lett.\ \textbf{60},
77 (1988).


\end{thebibliography}
\end{document}